\documentclass[twoside,11pt]{article}

\usepackage{jmlr2e}
\usepackage{amsmath}
\usepackage{amssymb}

\usepackage[english]{babel}
\usepackage{blindtext}
\usepackage{algorithm, algorithmicx, algpseudocode}
\usepackage{caption}
\usepackage{subcaption}
\usepackage{wrapfig}

\DeclareMathOperator*{\argmax}{argmax}

\jmlrheading{1}{}{}{}{}{}

\firstpageno{1}

\begin{document}

\title{Accelerating Black-Box Molecular Property Optimization by Adaptively Learning Sparse Subspaces}

\author{\name Farshud Sorourifar \email sorourifar.1@osu.edu\AND Thomas Banker \email thomas\_banker@berkeley.edu\AND Joel A. Paulson \email paulson.82@osu.edu  \\
\addr  Department of Chemical and Biomolecular Engineering, \\
       The Ohio State University, Columbus, OH 43210, USA}

\date{September 2023}

\maketitle
\vspace{-30pt}
\begin{abstract}
Molecular property optimization (MPO) problems are inherently challenging since they are formulated over discrete, unstructured spaces and the labeling process involves expensive simulations or experiments, which fundamentally limits the amount of available data. Bayesian optimization (BO) is a powerful and popular framework for efficient optimization of noisy, black-box objective functions (e.g., measured property values), thus is a potentially attractive framework for MPO. To apply BO to MPO problems, one must select a structured molecular representation that enables construction of a probabilistic surrogate model. Many molecular representations have been developed, however, they are all high-dimensional, which introduces important challenges in the BO process -- mainly because the curse of dimensionality makes it difficult to define and perform inference over a suitable class of surrogate models. This challenge has been recently addressed by learning a lower-dimensional encoding of a SMILE or graph representation of a molecule in an unsupervised manner and then performing BO in the encoded space. In this work, we show that such methods have a tendency to ``get stuck,'' which we hypothesize occurs since the mapping from the encoded space to property values is not necessarily well-modeled by a Gaussian process. We argue for an alternative approach that combines numerical molecular descriptors with a sparse axis-aligned Gaussian process model, which is capable of rapidly identifying sparse subspaces that are most relevant to modeling the unknown property function. We demonstrate that our proposed method substantially outperforms existing MPO methods on a variety of benchmark and real-world problems. Specifically, we show that our method can routinely find near-optimal molecules out of a set of more than $>100$k alternatives within 100 or fewer expensive queries.  
\end{abstract}

\section{Introduction}

Molecular property optimization (MPO) is the process of systematically improving the structural and/or functional properties of molecules to meet specific objectives. MPO is a critical step in a variety of scientific and engineering applications including chemistry, drug discovery, and material science. One can generally formulate MPO problems as follows
\begin{align} \label{eq:mpo}
    m^\star \in \argmax_{m \in \mathcal{M}} ~ F(m),
\end{align}
where $m$ is a molecule from the discrete set $\mathcal{M}$ and $F : \mathcal{M} \to \mathbb{R}$ is an unknown objective function that maps a molecule to a performance value.

The goal is thus to find the ``best'' molecule $m^\star$ that has the best performance $F(m^\star) \geq F(m)$ for all $m \in \mathcal{M}$ where $\mathcal{M}$ is a set with large but finite cardinality $| \mathcal{M} | < \infty$. This problem is easily solved if we could perfectly measure the property for every molecule; however, in reality, (i) we often only get noisy observations $y = F(m) + \varepsilon$ and (ii) the number of candidate molecules is very large (millions or more) such that we can only observe a relatively small number of options. In recent years, there has been a substantial amount of work on the use of machine learning as an effective tool to address these challenges \citep{bartok2017machine, von2020retrospective}. Most work on machine learning for MPO can be divided into two categories: \textit{guided search} and \textit{translation} \citep{hoffman2022optimizing}. In guided search, the idea is to construct predictive models over some type of molecular representation to sequentially select promising molecules for testing. 
Translation, on the other hand, treats the molecule generation problem as a sequence-to-sequence translation problem, which requires additional information that is not always available. Thus, in this work, we focus on developing a guided search approach.

To develop a guided search strategy, one must first select an effective numerical representation of the molecule. We can think of this representation as a function $R : \mathcal{M} \to \mathcal{X}$ that maps from molecule space $\mathcal{M}$ to a numerical feature space $\mathcal{X}$. As long as this mapping is invertible, we can equivalently express \eqref{eq:mpo} as finding $m^\star = R^{-1}(x^\star)$ where $x^\star \in \argmax_{x \in \mathcal{X}} f(x)$ and $f(x) = F \circ R^{-1}(x)$ is the objective as a function of the structured numerical representation vector $x$. The space $\mathcal{X}$ could be continuous or discrete and many different representations have been proposed in the literature including SMILES strings \citep{anderson1987smiles}, molecular graphs \citep{wieder2020compact}, molecular fingerprints \citep{cereto2015molecular}, and molecular descriptors \citep{mordred}. An important challenge with all of these representations is that they are naturally high-dimensional, which complicates the optimization problem. A recent line of work has looked to address this challenge by learning a lower-dimensional continuous latent representation in which one can more easily execute efficient search strategies such as Bayesian optimization (BO) \citep{frazier2018tutorial}. One of the most common examples is the combined use of a variational autoencoder and BO \citep{gomez2018automatic, griffiths2020constrained}. Let $z = E(x)$ denote an encoded latent representation of $x$. These methods proceed by using BO to maximize $g(z) = f(D(z))$ over $z$ (where a decoder transforms back to $x = D(z)$). If $E$ is learned using little-to-no property data, then there is no driving force for $g$ to have smoothness or continuity properties that would allow one to efficiently search over the latent space $\mathcal{Z}$ even if it has a smaller dimension than $\mathcal{X}$. Furthermore, it is possible that sparsity in the behavior of $f(x)$ is lost when transforming to $g(z)$. 

In this paper, we propose the \textbf{Mol}ecular \textbf{D}escriptors and \textbf{A}ctively \textbf{I}dentified \textbf{S}ubspaces (MolDAIS) framework, which is a new strategy to approach MPO problems that work directly in a numerical molecular feature space. In particular, we argue that molecular descriptors (i.e., outcomes of mathematical procedures applied to the symbolic representation of a molecule) \citep{MolecularDecriptorHandbook} combined with Gaussian process surrogate models defined on sparse axis-aligned subspaces (SAAS) provide an effective framework for MPO in the low-data regime. The motivation for our approach can be traced back to ideas in interpretable machine learning, which often posit the existence of a relatively small number of well-selected (understandable) features that can be used to accurately predict the desired property of interest. Since these key descriptors may not be known \textit{a priori}, one can attempt to learn them with sparse regression methods such as SISSO \citep{sisso}. These sparse regression methods, however, require that the target property depend linearly on the important descriptors in the feature set. They also do not directly capture uncertainty, which is crucial for navigating the exploration-exploitation tradeoff in BO. It turns out that we can address both of these challenges by taking advantage of the SAAS prior developed in \citep{saas}. Not only can SAAS sparsely pick out features from $x$,
the space is systematically updated as new information is collected, enabling adaptive learning of sparse and interpretable subspaces. The latter point significantly simplifies the inference task, which reduces the amount of data needed to make useful property predictions. We demonstrate the advantages of MolDAIS on three unique problems by comparing to existing MPO approaches. First, we consider a benchmark logP problem for which we can consistently find the best molecule out of $250$k candidates in only 100 iterations (substantially outperforming state-of-the-art alternatives). Second, we consider two real-world problems related to optimization over a class of organic molecules whose properties are computed from an expensive density function theory simulation.

\section{The MolDAIS Framework} \label{sec:moldais}

\subsection{Molecular Descriptors}

Molecular descriptors are defined as the \textit{``final result of a logical and mathematical procedure, which transforms chemical information encoded within a symbolic representation of a molecule into a useful number or the result of some standardized experiment''} \citep{MolecularDecriptorHandbook}. Many types of molecular descriptors have been developed.
Here, we focus on a set of more than 1800 descriptors that can be efficiently computed from the open-source Mordred software program \citep{mordred}. Examples of the descriptors include outcomes of rotationally and translationally invariant operations on a molecular graph as well as quantities relevant in chemistry (such as atomic weights). 
Using these descriptors, we can cast \eqref{eq:mpo} as the following equivalent optimization problem
\begin{align} \label{eq:mordred-opt}
    x^\star \in \argmax_{x \in \mathcal{X}} ~ f(x), 
\end{align}
where $f(x) = F(R^{-1}(x))$ denotes the performance value as a function of the Mordred representation $x = R(m)$ and $\mathcal{X} = [0,1]^D$ is the normalized numerical Mordred space defined over a $D$-dimensional hypercube ($D \sim 2000$ in this work).

\subsection{Bayesian Optimization with Sparse Axis-aligned Subspaces} \label{subsec:saasbo}

The challenge with solving \eqref{eq:mordred-opt} directly is that $\mathcal{X}$ is a high-dimensional space so it is difficult to apply efficient optimization strategies, such as Bayesian optimization (BO), due to the curse of dimensionality that manifests when attempting to build a surrogate model for $f(x)$ from initial data. In particular, Gaussian processes (GPs) are non-parametric function priors that are commonly used in BO due to their flexibility and natural uncertainty quantification abilities \citep{williams2006gaussian}. A GP over an input space $\mathcal{X}$ is fully specified by a prior mean function $\mu_0 : \mathcal{X} \to \mathbb{R}$ and prior covariance (or kernel) function $k : \mathcal{X} \times \mathcal{X} \to \mathbb{R}$. As commonly done, we will assume $\mu_0(x) = 0$ for all $x \in \mathcal{X}$, which can be achieved in practice by normalizing the data. The kernel function encodes information about the smoothness and rate of change of the unknown function, with a popular choice being the squared exponential (SE) kernel given by
\begin{align} \label{eq:sekernel}
    k^\psi( x, x' ) = \sigma_k^2 \exp\left\lbrace -\textstyle\frac{1}{2} \sum_{i=1}^D \rho_i (x_i - x'_i)^2 \right\rbrace,
\end{align}
where $\psi = \{ \rho_1, \ldots, \rho_D, \sigma_k^2 \}$ are the hyperparameters of the kernel that consist of inverse lengthscales $\{ \rho_i \}_{i=1}^D$ for each dimension and the output scale $\sigma_k^2$. For fixed $\psi$, the posterior distribution $p(f(x_\star) | \mathcal{D}_n) \sim \mathcal{N}( \mu_n(x_\star), \sigma_n^2(x_\star) )$ at a test point $x_\star$ given past observations $\mathcal{D}_n = \{ (x_i, y_i) \}_{i=1}^n$ can be analytically computed. Given this predictive distribution, standard BO proceeds by selecting the next evaluation point $x_{n+1} \in \argmax_{x \in \mathcal{X}} \alpha_n(x)$ that maximizes an acquisition function $\alpha_n : \mathcal{X} \to \mathbb{R}$, which should be chosen to provide a good measure of the potential benefit of querying $f$ at every $x \in \mathcal{X}$ in the future. By defining our utility function as the best observed sample $u(\mathcal{D}_n) = \max_{(x,y) \in \mathcal{D}_n} y_n$, we can derive the \textit{expected improvement} (EI) acquisition function as the expected increase in utility $\alpha_n(x) = \text{EI}_n(x | \eta, \psi) = \mathbb{E}_n\{ u(\mathcal{D}_{n+1}) - u(\mathcal{D}_n) \}$ where $\eta = \max_n y_n$ is the best observed value so far. 
For a standard GP model, EI has a simple closed-form solution \cite{jones1998efficient}. 

The main challenge with the standard BO strategy for \eqref{eq:mordred-opt} is that, when $D$ is large, the space of possible functions mapping $\mathcal{X}$ to $\mathbb{R}$ is too large to learn even assuming some degree of smoothness imparted by the SE kernel \eqref{eq:sekernel}. Without additional prior information, a natural way to deal with this challenge is to assume a hierarchy of relevance in the dimensions such that we can focus on the subset of features in $x = \{ x_1, x_2, \ldots, x_D \}$ that are important for our property of interest. We choose to exploit the sparse axis-aligned subspace (SAAS) prior to accomplish this task. In short, SAAS is a prior over the kernel hyperparameters that induces sparse structure in the inverse squared lengthscales $\rho_i$. Small values of $\rho_i$ imply dimension $i$ is \textit{unimportant} in the prediction, so a half-Cauchy prior is used to concentrate their values near zero. Only once observations provide enough evidence that dimension $i$ is important can $\rho_i$ escape zero. As more data is accumulated, the number of dimensions allowed to escape increases, allowing more $\rho_i$ to be ``activated'', leading to a richer class of functions. Interested readers are referred to \cite{saas} for details on the SAAS prior as well as the training procedure that exploits an established Hamiltonian Monte Carlo method to generate $L$ approximate posterior samples for the kernel hyperparameters $\{ \psi_l \}_{l=1}^L$. Therefore, given $n$ previous property evaluations for different molecules, the next molecule we want to sample is selected according to $m_{n+1} = R^{-1}(x_{n+1})$
\begin{align} \label{eq:saasbo}
    x_{n+1} \in \argmax_{x \in \mathcal{X}} ~ \textstyle \frac{1}{L} \sum_{l=1}^L \text{EI}_n(x | \eta, \psi_l).
\end{align}
These steps have been implemented in the BoTorch package \citep{balandat2020botorch} for which \eqref{eq:saasbo} can be tackled using efficient gradient-based optimization methods.

\section{Experiments}

\subsection{Baseline Methods}

We compare MolDAIS to several baselines that are a mixture of different choices of the starting molecular feature space, the learned latent representation, and type of BO method. All methods are provided with 10 randomly chosen initial samples and a budget of 90 additional samples and results are shown for 5 independent replications of each algorithm.

\textbf{LADDER} is a recently proposed MPO method that combines molecular fingerprints (FP) with a junction-tree autoencoder (JTAE) to construct a latent representation of the FP space ~\cite{ladder}. A standard BO method is then used to actively select new samples in the latent space. We use the default settings from the original paper

\textbf{SAAS-FP-JTAE} is a variant of LADDER that uses the same latent representation (learned by applying a JTAE to molecular FPs) but now uses the SAASBO strategy described in Section \ref{subsec:saasbo}. This is meant to study the impact of the starting molecular descriptor space, as we expect the SAAS prior to be less effective in the encoded latent space. 

\textbf{SBO} refers to the standard BO method that optimizes directly over the high-dimensional molecular descriptor $\mathcal{X}$ space using the EI acquisition function. This is meant to study the importance of the SAAS prior for accelerating convergence.

\textbf{SBO-PCA} is a slight modification to SBO that replaces the high-dimensional $\mathcal{X}$ space with a new lower-dimensional latent space constructed by applying principal component analysis (PCA) to the unlabeled molecular descriptor data. The size of the latent space is chosen such that 99\% of the variance in the $\mathcal{X}$ data is captured.

\textbf{SLR} refers to sequential linear regression, which is a very naive version of MolDAIS that uses sparse linear regression to identify a model with a small number of non-zero coefficients. This linear surrogate model is then maximized directly to select the next sample. 

\textbf{Random} refers to a standard random search strategy wherein the next sample is chosen uniformly at random in $\mathcal{X}$. This method, which does not exploit any past data, is meant to provide a lower bound on performance. 

\subsection{Maximizing LogP over the Zinc Molecule Dataset}
We first consider the problem of finding molecules with the best drug-like properties \cite{kusner2017grammar}. In particular, the goal is to maximize the water-octanol partition coefficient (logP) over the space of molecules. As done in previous work, we consider the Zinc molecule data set that consists of $250,000$ commercially-available molecules. 

The results are shown in Figure \ref{fig:results} (Left), which shows that MolDAIS clearly outperforms all other methods, as it is able to find the highest logP value by sampling less than 0.04\% of the candidates in all replicates. This represents a more than 30\% improvement over the best candidates found with all other methods under the same conditions. Even a reduced space variant of MolDAIS, which removes any feature suspected to highly correlate to logP, still outperforms all other tested methods. Interestingly, LADDER performs relatively similar to the other methods on average and actually results in a larger distribution of outcomes.

\subsection{Maximizing Solvation Free Energy for Quinone Molecule Class} 
\label{subsec:max-Gsolv}

Next, we consider the problem of finding the best molecule from the quinone class ~\cite{Tabor2019} that has the largest Gibbs free energy of solvation $\Delta G_\text{solv}$, which is an important property for battery materials and pharmaceuticals. Although one can compute $\Delta G_\text{solv}$ using density functional theory (DFT), it requires two separate simulations in different phases, which takes around 24 CPU hours per molecule on a supercomputing cluster. We consider a space of $>100,000$ quinone molecules and look to maximize $\Delta G_\text{solv}$. The results are shown in Figure \ref{fig:results} (Middle), which again outperform all considered alternatives. MolDAIS finds the global solution within around 10 iterations. Since the latent space built with LADDER may not be compatible with the quinone class, we do not directly compare against LADDER or SAAS-FP-JTAE in this example. To gain more insight into why MolDAIS can achieve such strong results, we perform additional analysis on the SAAS-GP model (Figure \ref{fig:results}; Right). The top plot shows the $\rho_i$ values sorted in decreasing order; we see a clear separation in terms of the top 4 values and the remaining values, indicating a small number of Mordred descriptors are needed to predict $\Delta G_\text{solv}$. Furthermore, the bottom plot shows the root mean squared error (RMSE) on a held-out set of 100 test molecules based on randomly sampled training sets of different sizes. We see that the RMSE values are significantly smaller even when only 20 data points are known, indicating the SAAS prior does enable learning a model structure with improved prediction accuracy.
See Appendix \ref{sec:E0_casestudy} for similar results achieved on a reduction potential maximization problem. 
    
\iftrue
\begin{figure}[tb!]
    \centering
    \includegraphics[width=\textwidth]{ 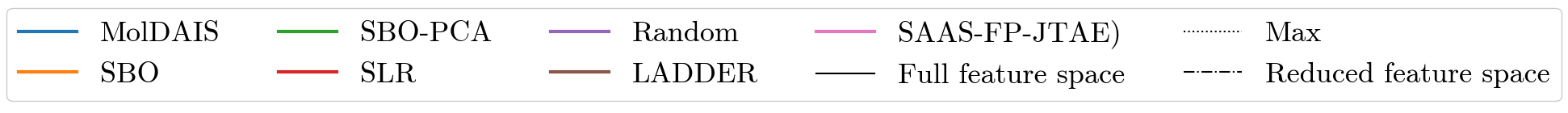}
    \\
    \includegraphics[width=.32\textwidth]{ 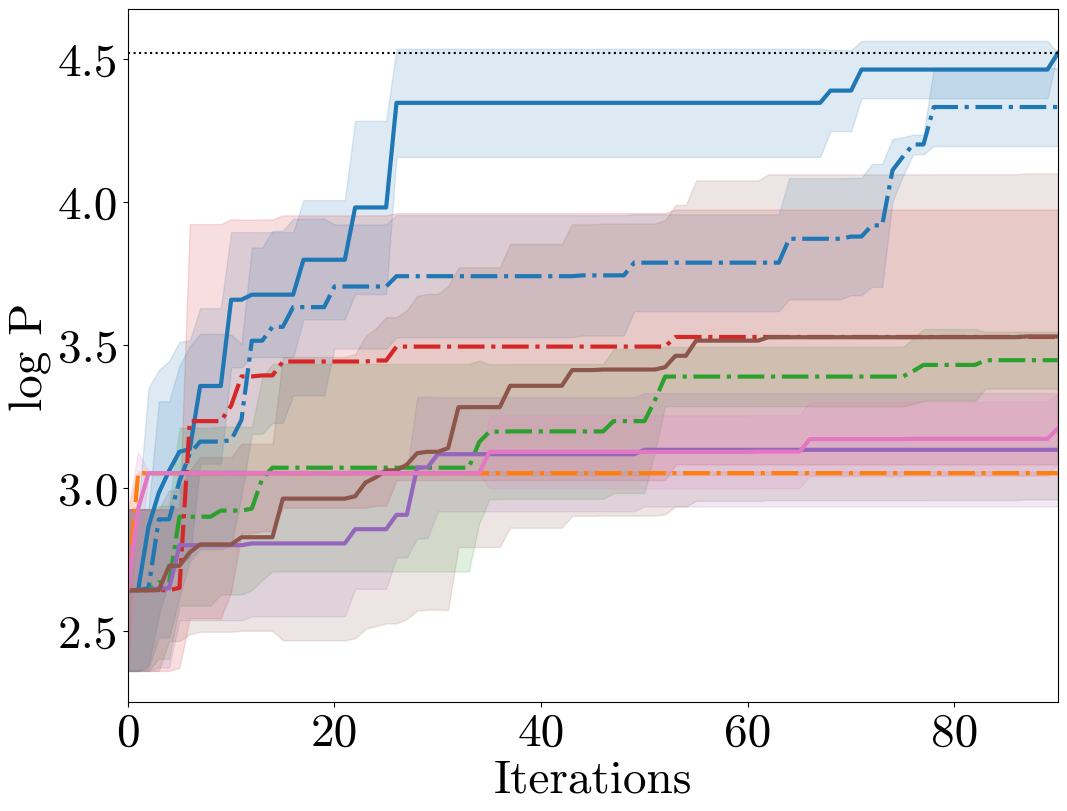}
    \includegraphics[width=.32\textwidth]{ 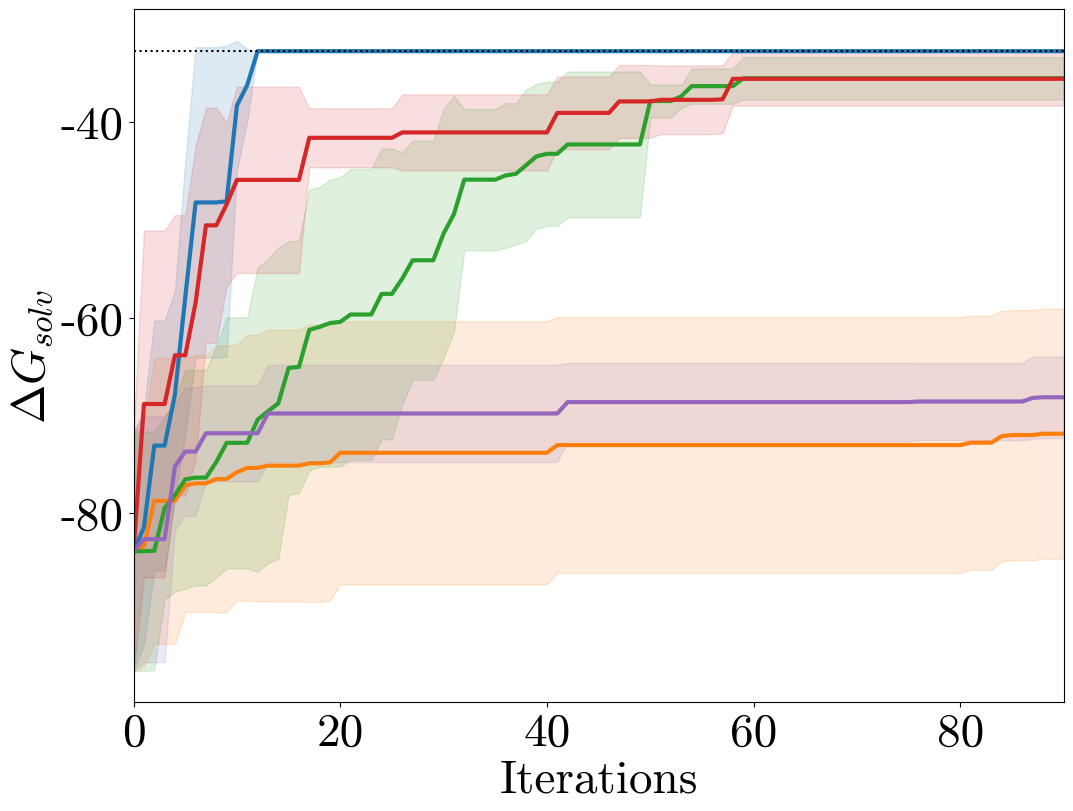} 
    \includegraphics[width=.32\textwidth]{ 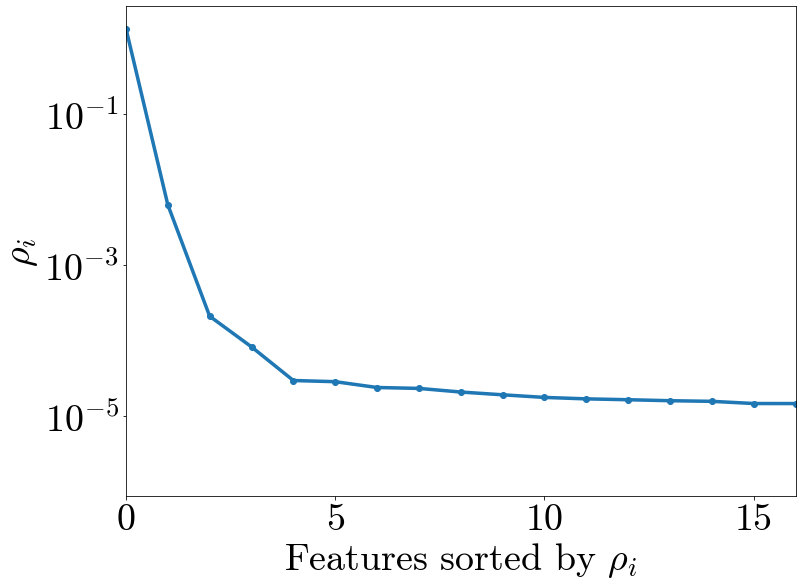} 
    \\
    \includegraphics[width=.32\textwidth]{ 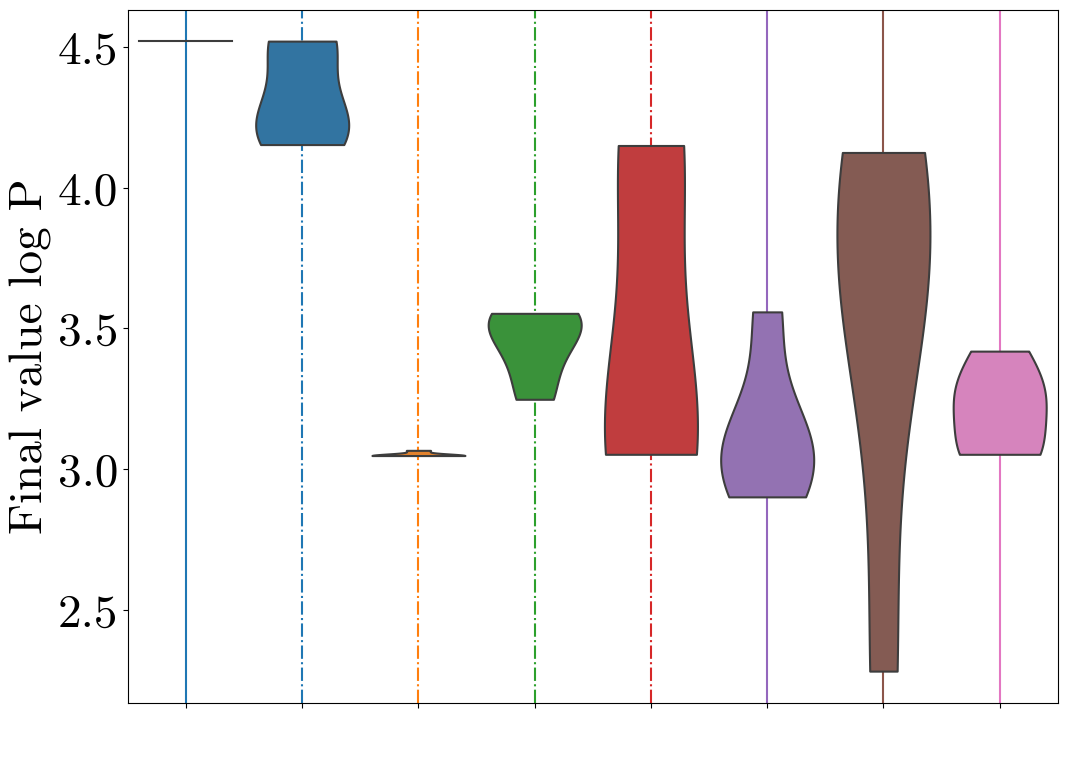}
    \includegraphics[width=.32\textwidth]{ 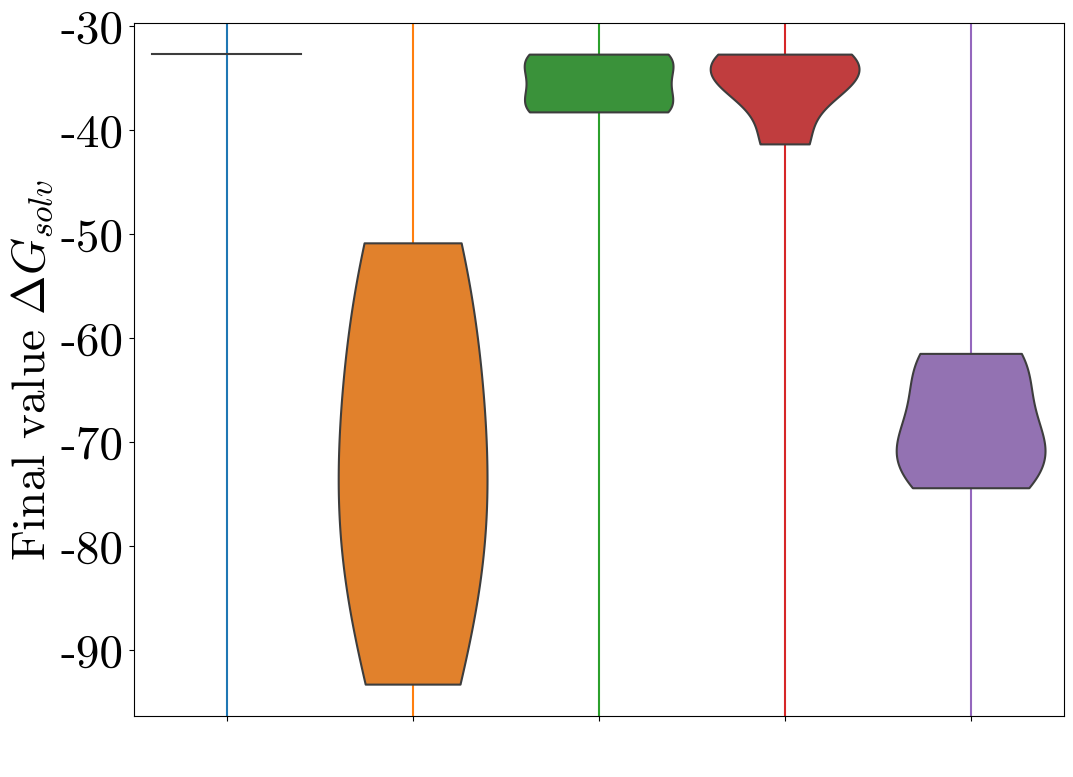}   
    \includegraphics[width=.32\textwidth]{ 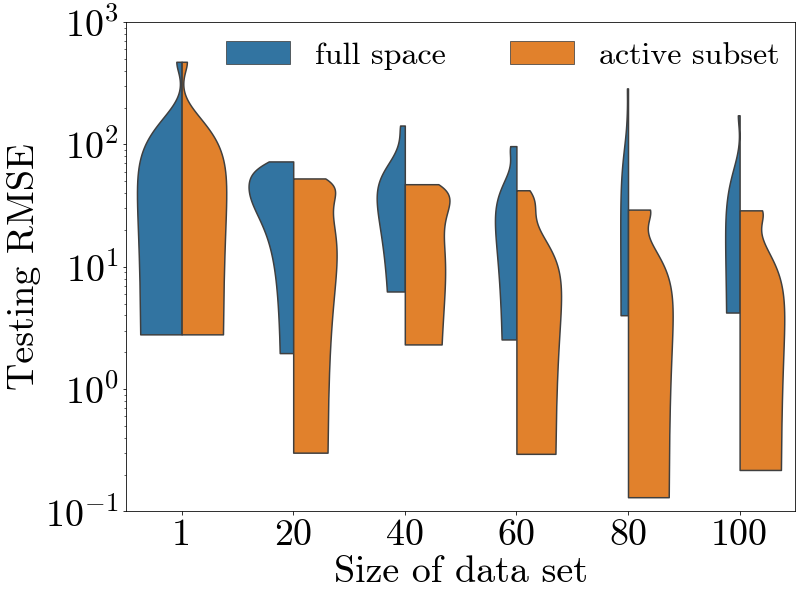}     
    \caption{\textbf{Left:} logP maximization. \textbf{Middle:} Solvation free energy maximization. \textbf{Right:} Feature importance and test accuracy for solvation free energy GP model. 
    }
    \label{fig:results}
\end{figure}
\fi

\section{Conclusions}

In this work, we propose a new molecular property optimization (MPO) method, MolDAIS, that can efficiently identify high-performance molecules in the low-data regime. MolDAIS combines molecular descriptors with a sparse axis-aligned subspace (SAAS) prior to adaptively learn sparse and interpretable subsets of the high-dimensional molecular feature space that can be used within a Bayesian optimization (BO) framework to directly balance exploration and exploitation of the search space. We empirically show that MolDAIS outperforms existing MPO methods on benchmark and real-world problems. Furthermore, in some cases, MolDAIS can find near globally optimal molecules with 100 or less queries out of more than 100k candidates without any prior information (i.e., in a fully black-box manner). 

\clearpage
\bibliographystyle{IEEEannot}

\bibliography{refrences}

\begin{thebibliography}{19}
\providecommand{\natexlab}[1]{#1}
\providecommand{\url}[1]{\texttt{#1}}
\expandafter\ifx\csname urlstyle\endcsname\relax
  \providecommand{\doi}[1]{doi: #1}\else
  \providecommand{\doi}{doi: \begingroup \urlstyle{rm}\Url}\fi

\bibitem[Anderson et~al.(1987)Anderson, Veith, and
  Weininger]{anderson1987smiles}
Eric Anderson, Gilman~D Veith, and David Weininger.
\newblock \emph{{SMILES}, a line notation and computerized interpreter for
  chemical structures}.
\newblock US Environmental Protection Agency, Environmental Research
  Laboratory, 1987.

\bibitem[Balandat et~al.(2020)Balandat, Karrer, Jiang, Daulton, Letham, Wilson,
  and Bakshy]{balandat2020botorch}
Maximilian Balandat, Brian Karrer, Daniel Jiang, Samuel Daulton, Ben Letham,
  Andrew~G Wilson, and Eytan Bakshy.
\newblock {BoTorch: A framework for efficient Monte-Carlo Bayesian
  optimization}.
\newblock \emph{Advances in Neural Information Processing Systems},
  33:\penalty0 21524--21538, 2020.

\bibitem[Bart{\'o}k et~al.(2017)Bart{\'o}k, De, Poelking, Bernstein, Kermode,
  Cs{\'a}nyi, and Ceriotti]{bartok2017machine}
Albert~P Bart{\'o}k, Sandip De, Carl Poelking, Noam Bernstein, James~R Kermode,
  G{\'a}bor Cs{\'a}nyi, and Michele Ceriotti.
\newblock Machine learning unifies the modeling of materials and molecules.
\newblock \emph{Science Advances}, 3\penalty0 (12):\penalty0 e1701816, 2017.

\bibitem[Cereto-Massagu{\'e} et~al.(2015)Cereto-Massagu{\'e}, Ojeda, Valls,
  Mulero, Garcia-Vallv{\'e}, and Pujadas]{cereto2015molecular}
Adri{\`a} Cereto-Massagu{\'e}, Mar{\'\i}a~Jos{\'e} Ojeda, Cristina Valls,
  Miquel Mulero, Santiago Garcia-Vallv{\'e}, and Gerard Pujadas.
\newblock Molecular fingerprint similarity search in virtual screening.
\newblock \emph{Methods}, 71:\penalty0 58--63, 2015.

\bibitem[Deshwal and Doppa(2021)]{ladder}
Aryan Deshwal and Jana Doppa.
\newblock Combining latent space and structured kernels for bayesian
  optimization over combinatorial spaces.
\newblock In M.~Ranzato, A.~Beygelzimer, Y.~Dauphin, P.S. Liang, and J.~Wortman
  Vaughan, editors, \emph{Advances in Neural Information Processing Systems},
  volume~34, pages 8185--8200. Curran Associates, Inc., 2021.
\newblock URL
  \url{https://proceedings.neurips.cc/paper_files/paper/2021/file/44e76e99b5e194377e955b13fb12f630-Paper.pdf}.

\bibitem[Eriksson and Jankowiak(2021)]{saas}
David Eriksson and Martin Jankowiak.
\newblock High-dimensional bayesian optimization with sparse axis-aligned
  subspaces.
\newblock In \emph{Uncertainty in Artificial Intelligence}, pages 493--503.
  PMLR, 2021.

\bibitem[Frazier(2018)]{frazier2018tutorial}
Peter~I. Frazier.
\newblock A tutorial on bayesian optimization, 2018.

\bibitem[G{\'o}mez-Bombarelli et~al.(2018)G{\'o}mez-Bombarelli, Wei, Duvenaud,
  Hern{\'a}ndez-Lobato, S{\'a}nchez-Lengeling, Sheberla, Aguilera-Iparraguirre,
  Hirzel, Adams, and Aspuru-Guzik]{gomez2018automatic}
Rafael G{\'o}mez-Bombarelli, Jennifer~N Wei, David Duvenaud, Jos{\'e}~Miguel
  Hern{\'a}ndez-Lobato, Benjam{\'\i}n S{\'a}nchez-Lengeling, Dennis Sheberla,
  Jorge Aguilera-Iparraguirre, Timothy~D Hirzel, Ryan~P Adams, and Al{\'a}n
  Aspuru-Guzik.
\newblock Automatic chemical design using a data-driven continuous
  representation of molecules.
\newblock \emph{ACS Central Science}, 4\penalty0 (2):\penalty0 268--276, 2018.

\bibitem[Griffiths and Hern{\'a}ndez-Lobato(2020)]{griffiths2020constrained}
Ryan-Rhys Griffiths and Jos{\'e}~Miguel Hern{\'a}ndez-Lobato.
\newblock Constrained {B}ayesian optimization for automatic chemical design
  using variational autoencoders.
\newblock \emph{Chemical Science}, 11\penalty0 (2):\penalty0 577--586, 2020.

\bibitem[Hoffman et~al.(2022)Hoffman, Chenthamarakshan, Wadhawan, Chen, and
  Das]{hoffman2022optimizing}
Samuel~C Hoffman, Vijil Chenthamarakshan, Kahini Wadhawan, Pin-Yu Chen, and
  Payel Das.
\newblock Optimizing molecules using efficient queries from property
  evaluations.
\newblock \emph{Nature Machine Intelligence}, 4\penalty0 (1):\penalty0 21--31,
  2022.

\bibitem[Jones et~al.(1998)Jones, Schonlau, and Welch]{jones1998efficient}
Donald~R Jones, Matthias Schonlau, and William~J Welch.
\newblock Efficient global optimization of expensive black-box functions.
\newblock \emph{Journal of Global Optimization}, 13:\penalty0 455--492, 1998.

\bibitem[Kusner et~al.(2017)Kusner, Paige, and
  Hernandez-Lobato]{kusner2017grammar}
Matt~J Kusner, Brooks Paige, and Jose~Miguel Hernandez-Lobato.
\newblock Grammar variational autoencoder.
\newblock \emph{International Conference on Machine Learning}, pages
  1945--1954, 2017.

\bibitem[Moriwaki et~al.(2018)Moriwaki, Tian, Kawashita, and Takagi]{mordred}
Hirotomo Moriwaki, Yu-Shi Tian, Norihito Kawashita, and Tatsuya Takagi.
\newblock Mordred: a molecular descriptor calculator.
\newblock \emph{Journal of Cheminformatics}, 10\penalty0 (1):\penalty0 4, Feb
  2018.
\newblock ISSN 1758-2946.
\newblock \doi{10.1186/s13321-018-0258-y}.
\newblock URL \url{https://doi.org/10.1186/s13321-018-0258-y}.

\bibitem[Ouyang et~al.(2018)Ouyang, Curtarolo, Ahmetcik, Scheffler, and
  Ghiringhelli]{sisso}
Runhai Ouyang, Stefano Curtarolo, Emre Ahmetcik, Matthias Scheffler, and
  Luca~M. Ghiringhelli.
\newblock Sisso: A compressed-sensing method for identifying the best
  low-dimensional descriptor in an immensity of offered candidates.
\newblock \emph{Phys. Rev. Mater.}, 2:\penalty0 083802, Aug 2018.
\newblock \doi{10.1103/PhysRevMaterials.2.083802}.
\newblock URL \url{https://link.aps.org/doi/10.1103/PhysRevMaterials.2.083802}.

\bibitem[Tabor et~al.(2019)Tabor, Gómez-Bombarelli, Tong, Gordon, Aziz, and
  Aspuru-Guzik]{Tabor2019}
Daniel~P. Tabor, Rafael Gómez-Bombarelli, Liuchuan Tong, Roy~G. Gordon,
  Michael~J. Aziz, and Alán Aspuru-Guzik.
\newblock Mapping the frontiers of quinone stability in aqueous media:
  implications for organic aqueous redox flow batteries.
\newblock \emph{J. Mater. Chem. A}, 7:\penalty0 12833--12841, 2019.
\newblock \doi{10.1039/C9TA03219C}.
\newblock URL \url{http://dx.doi.org/10.1039/C9TA03219C}.

\bibitem[Todeschini and Consonni(2000)]{MolecularDecriptorHandbook}
R.~Todeschini and V~Consonni.
\newblock \emph{Frontmatter}, pages i--xxi.
\newblock John Wiley \& Sons, Ltd, 2000.
\newblock ISBN 9783527613106.
\newblock \doi{https://doi.org/10.1002/9783527613106.fmatter}.
\newblock URL
  \url{https://onlinelibrary.wiley.com/doi/abs/10.1002/9783527613106.fmatter}.

\bibitem[von Lilienfeld and Burke(2020)]{von2020retrospective}
O~Anatole von Lilienfeld and Kieron Burke.
\newblock Retrospective on a decade of machine learning for chemical discovery.
\newblock \emph{Nature Communications}, 11\penalty0 (1):\penalty0 4895, 2020.

\bibitem[Wieder et~al.(2020)Wieder, Kohlbacher, Kuenemann, Garon, Ducrot,
  Seidel, and Langer]{wieder2020compact}
Oliver Wieder, Stefan Kohlbacher, M{\'e}laine Kuenemann, Arthur Garon, Pierre
  Ducrot, Thomas Seidel, and Thierry Langer.
\newblock A compact review of molecular property prediction with graph neural
  networks.
\newblock \emph{Drug Discovery Today: Technologies}, 37:\penalty0 1--12, 2020.

\bibitem[Williams and Rasmussen(2006)]{williams2006gaussian}
Christopher~KI Williams and Carl~Edward Rasmussen.
\newblock Gaussian processes for machine learning, 2006.

\end{thebibliography}

\newpage

\appendix

\section{Maximizing Reduction Potential} \label{sec:E0_casestudy}

\begin{figure}[h!]
    \centering
    \includegraphics[width=.8\textwidth]{ 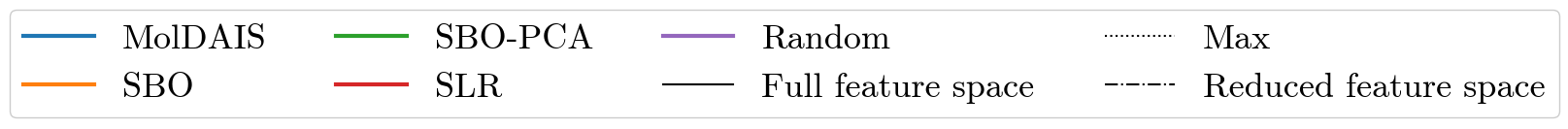}\\
    \includegraphics[width=.45\textwidth]{ 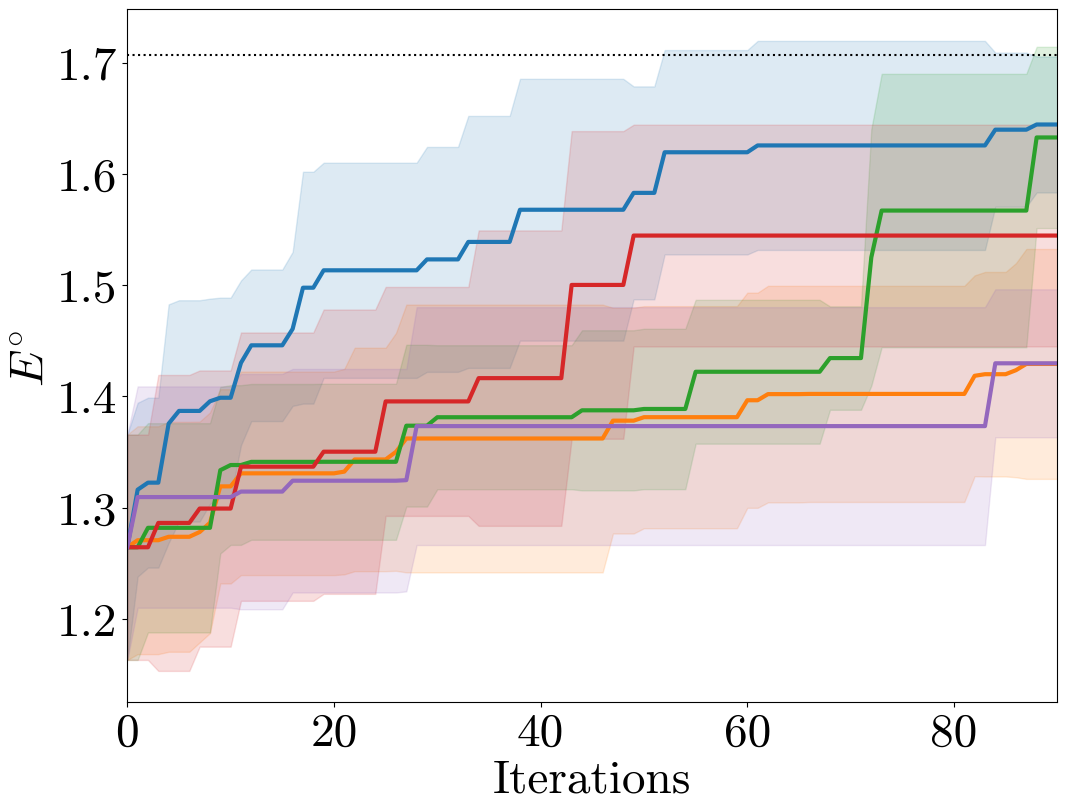}
    \includegraphics[width=.45\textwidth]{ 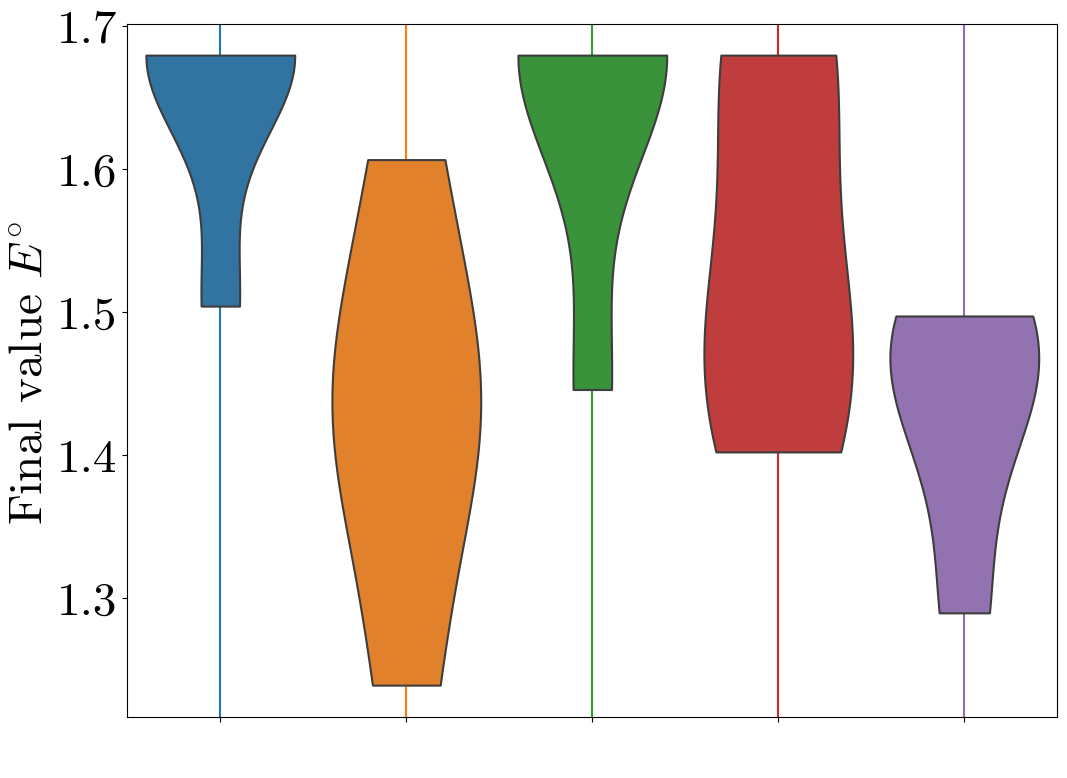}\\
    \includegraphics[width=.45\textwidth]{ 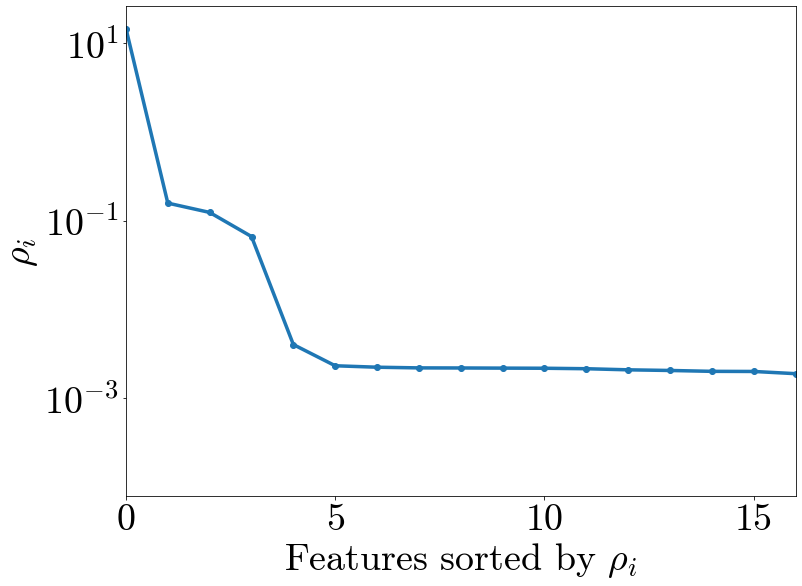}
    \includegraphics[width=.45\textwidth]{ 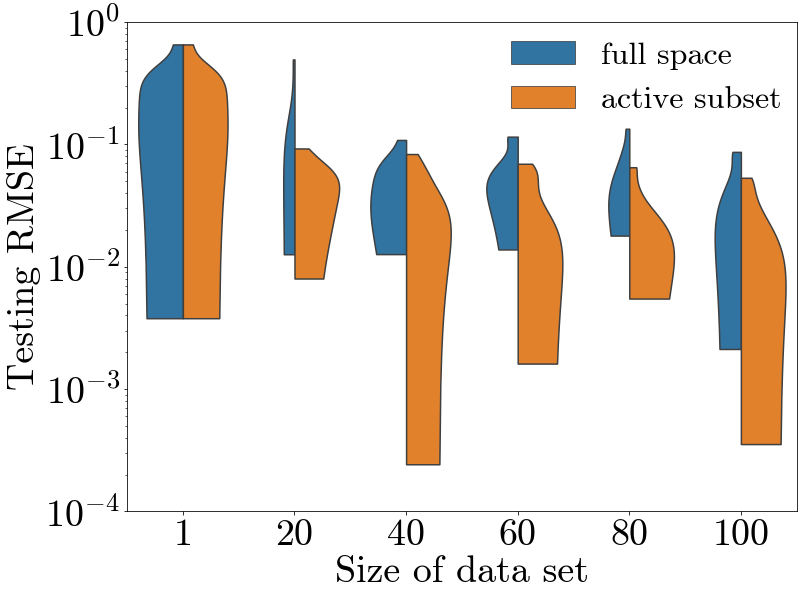}    
    \caption{\textbf{Top Left:} Convergence results for maximization of redox potential versus number of iterations. \textbf{Top Right:} The distribution in the best maximum found at the final iteration over the replicates. \textbf{Bottom Left:} Inverse squared lengthscale values sorted in descending order for the SAAS-GP model at the final iteration. \textbf{Bottom Right:} Test RMSE values for the standard GP and SAAS-GP models given random training sets of different sizes. 
    }\label{fig:eo-results}
\end{figure}

To further illustrate the flexibility of the proposed MolDAIS method, we consider another variant of an MPO problem defined over the quinone molecule class considered in Section \ref{subsec:max-Gsolv}. Specifically, we now consider the reduction (or redox) potential that are a measure of energy required to reduce or oxide a molecule relative to the standard hydrogen electrode, which can again be predicted using DFT. The results for maximizing redox potential $E^0$ are shown in Figure \ref{fig:eo-results}. Even though this is a completely different property than $\Delta G_\text{solv}$, MolDAIS still outperforms all considered alternative methods, though the SBO-PCA method does perform fairly close towards the final iterations. We again see a clear separation in terms of the number of relevant dimensions (those with relatively large $\rho_i$ values), though the overall values are higher than in the $\Delta G_\text{solv}$ that is likely the source of slightly worse performance. Similarly, from a prediction quality point-of-view, the SAAS-GP still outperforms the traditional GP, though there is less of a gap than in the previous cases.

\end{document}